\documentstyle[epsfig]{mn}
\begin{document}

\title[] { Jet-dominated states: an alternative to advection across
black hole event horizons in `quiescent' X-ray binaries
}
\author[R.~P.~Fender et al.] {
R. P. Fender$^1$, E. Gallo$^1$, P. G. Jonker$^2$\\
$^1$Astronomical Institute `Anton Pannekoek', University of Amsterdam, 
Kruislaan 403, 1098 SJ Amsterdam, The Netherlands\\
$^2$ Institute of Astronomy, Madingley Road, Cambridge CB3 0HA\\}

\maketitle

\begin{abstract}

We demonstrate that at relatively low mass accretion rates, black hole
candidate (BHC) X-ray binaries (XRBs) should enter `jet-dominated'
states, in which the majority of the liberated accretion power is in
the form of a (radiatively inefficient) jet and not dissipated as
X-rays in the accretion flow. This result follows from the empirically
established non-linear relation between radio and X-ray power from
low/hard state BHC XRBs, which we assume also to hold for neutron star
(NS) XRBs.  Conservative estimates of the jet power indicate that all
BHC XRBs in `quiescence' should be in this jet-dominated regime.  In
combination with an additional empirical result, namely that BHC XRBs
are more `radio loud' than NS XRBs, we find that in quiescence NS XRBs
should be up to two orders of magnitude more luminous in X-rays than
BHC XRBs, without requiring any significant advection of energy into a
black hole.  This ratio is as observed, and such observations should
therefore no longer be considered as direct evidence for the existence
of black hole event horizons. Furthermore, even if BHCs do contain
black holes with event horizons, this work demonstrates that there is
no requirement for the advection of significant amounts of accretion
energy across the horizon.

\end{abstract}

\begin{keywords}

binaries: close -- radio continuum: stars -- ISM:jets and outflows --
stars: neutron -- black hole physics

\end{keywords}


\section{Introduction}

Proving the existence of black holes remains a key goal of
observational high energy astrophysics. While dynamical evidence
(e.g. Charles 1998) convincingly demonstrates the existence of compact
accreting objects in binary systems which have masses in excess of the
highest theoretical limit for a neutron star ($\sim 3 M_{\odot}$), and
are therefore strong black hole candidates (BHCs), we cannot rule out
the possibility that some as-yet-unconsidered state of matter may
provide an alternative explanation.

As an alternative approach, in recent years much attention has been
focussed on finding evidence for black hole event horizons. One
promising and actively pursued route has been a comparison of the
X-ray luminosities of BHC and neutron star (NS) X-ray binaries (XRBs)
in `quiescence'. In such states, black hole accretion could be
advection dominated and considerably fainter than neutron stars. This
is indeed what has been found observationally, with `quiescent' BHCs
being typically two to three orders of magnitude (in Eddington units)
less luminous than their NS XRB equivalents.  This has been claimed to
represent some of the strongest evidence to date for the existence of
black hole event horizons (Narayan, Garcia \& McClintock 1997; Menou
et al. 1999; Garcia et al. 2001). However, alternatives to this
interpretation have also been discussed (Campana \& Stella 2000;
Bildsten \& Rutledge 2000; Abramowicz, Kluzniak \& Lasota 2002).
Abramowicz et al. (2002), in particular, stress that `absence of
evidence is not evidence of absence', and draw attention to
alternatives to black holes. Even if BHCs {\em do} contain black hole
with event horizons, it is important to establish how much, if any, of the
potential accretion energy may be being advected across their
horizons.

In a series of important and related observations, in recent years it
has been established that jets are an integral and relatively
ubiquitous component of the process of accretion in both black hole
and neutron star X-ray binaries (e.g. Mirabel \& Rodriguez 1999;
Fender 2002). We are now beginning to understand just how powerful
these jets may be. Corbel et al. (2003) discovered that, over four
orders of magnitude in X-ray luminosity, the relation between radio
and X-ray luminosity for the BHC X-ray binary GX 339-4 has the form
$L_{\rm radio} \propto L_{\rm X}^b$, where $b = 0.706 \pm 0.011$ for
$L_{\rm X}$ in the 3--9 keV range ($b$ increases slightly with the
increasing energy of the X-ray band used for comparison).  Gallo,
Fender \& Pooley (2003) demonstrated that the same relation holds over
a comparable range in X-ray luminosity for the transient V404 Cyg (GS
2023+338), and furthermore that the data for all measured low/hard
state sources is consistent with a such a Universal relation holding
for all of them. This power law relation between radio and X-ray
luminosity is a key observational discovery providing clues to the
underlying physics of the disc--jet coupling.


Are there differences in jet power between the BHCs and NS XRBs ?
Fender \& Kuulkers (2001) found that BHC XRBs were, in general between
10--100 times more `radio loud' (in the sense of the radio to soft
X-ray ratio) than neutron star binaries. Migliari et al. (2003)
compared the radio strength of the atoll-type neutron star binary 4U
1728-34 with the comparable state and X-ray luminosity of BHCs, and found a
ratio of radio loudness $R_{\rm radio} = (L_{\rm radio} / L_{\rm
X})_{\rm BH} / (L_{\rm radio} / L_{\rm X})_{\rm ns} \sim 30$.  The
origin of this difference in radio loudness is not clear (see Fender
\& Kuulkers 2001 for a discussion).


\begin{figure*}
\centerline{{\epsfig{file=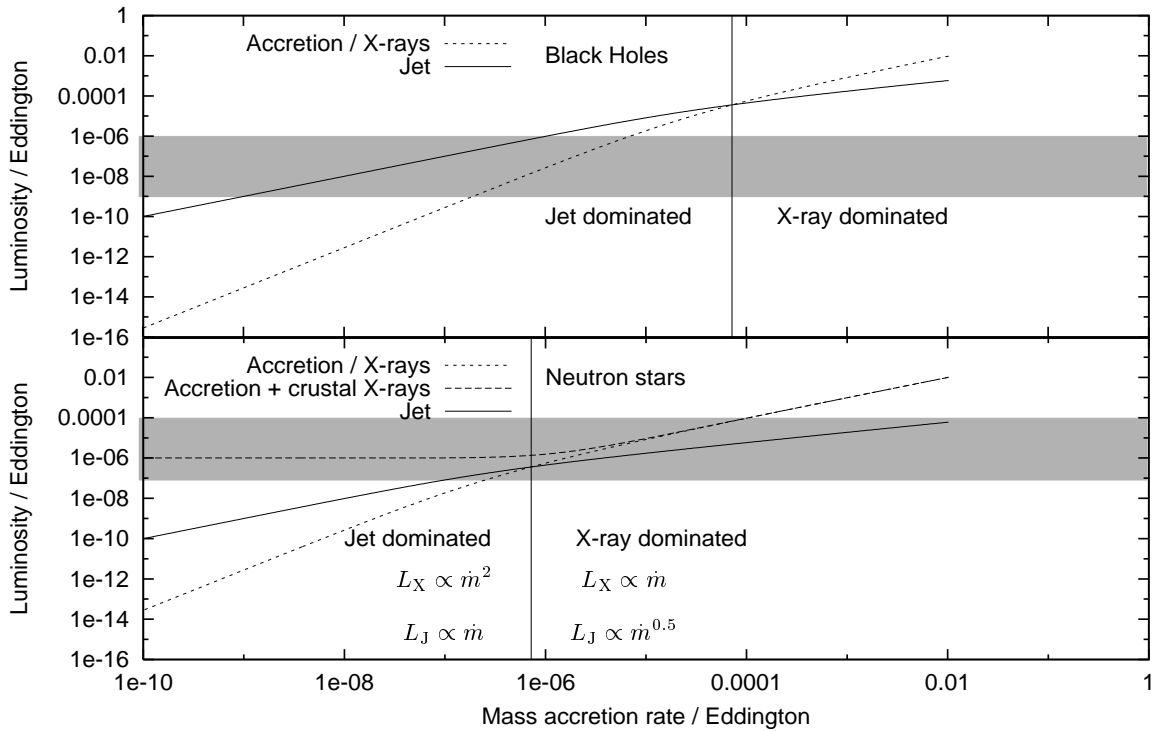, width=16cm, height=11cm, angle=0}}}
\caption{Variation of X-ray luminosity and jet power as a function of
  mass accretion rate, in our model, for neutron star and black hole X-ray
  binaries. Two regimes exist, `X-ray dominated' and `Jet dominated',
  with the transition from the former to the latter occurring at two
  orders of magnitude lower accretion rate in neutron stars than in
  black holes, due to their lower `radio loudness'. In the `X-ray'
  dominated regime, $L_{\rm X} \propto \dot{m}$, but in the `Jet
  dominated' regime $L_{\rm X} \propto \dot{m}^2$. The transition
  between the two regimes occurs at an X-ray luminosity $L_{\rm X} =
  A^2$, where $A_{\rm BH} \sim 6\times 10^{-3}$ and $A_{\rm NS} \sim
  6\times 10^{-4}$. The shaded areas indicate the range of X-ray
  luminosities observed in `quiescence' from the two types of X-ray
  binary.  If this model is correct, {\em all} of the quiescent black
  hole binaries are in the jet-dominated regime.}
\end{figure*}

\section{Jet-dominated states in black hole candidates}

In the following all luminosities and accretion rates are in Eddington
units, where the Eddington luminosity is $\sim 1.3 \times 10^{38}
(M/M_{\odot})$ erg s$^{-1}$, where $M$ is the mass of the accreting
compact star. The Eddington accretion rate, defined as that
accretion rate at which the Eddington luminosity is achieved, is, for
an accretion efficiency of $\sim 10$\% (ie. $\sim 0.1 \dot{m}c^2$ is
liberated during the accretion process) approximately $1.4 \times
10^{18} (M/M_{\odot})$ g s$^{-1}$. 

We assume that the total power output $L_{\rm total}$ from an
X-ray binary in a `low/hard' or analogous state is a combination of
the radiative luminosity of the flow ($L_{\rm X}$, directly observed
as X-rays) and jet power ($L_{\rm J}$ indirectly traced by e.g. radio
flux density):

\begin{equation}
L_{\rm total} = L_{\rm X} + L_{\rm J}
\end{equation}

Now we already know (Corbel et al. 2002; Gallo et al. 2003) the
relation between radio ($L_{\rm radio}$) and X-ray luminosity:

\begin{equation}
L_{\rm radio} \propto L_{\rm X}^{0.7}
\end{equation}


How does observed radio flux relate to jet power; i.e. what is the
relation between $L_{\rm radio}$ and $L_{\rm J}$ ? In models of
optically thick jets (e.g. Blandford \& K\"onigl 1979; Falcke \&
Biermann 1996; Markoff, Falcke \& Fender 2001; Heinz \& Sunyaev 2003),
the following scaling applies:

\begin{equation}
L_{\rm radio} \propto L_{\rm J}^{1.4}
\label{jetradio}
\end{equation}

%
%
Combining equations (2) and (3):

\begin{equation}
L_{\rm J} \propto L_{\rm X}^{0.5}
\end{equation}

therefore

\begin{equation}
L_{\rm total} = L_{\rm X} + A L_{\rm X}^{0.5}
\end{equation}

which provides the relation between total power and X-ray luminosity.
The normalisation $A$ between can be estimated. Fender (2001) and
Corbel \& Fender (2002) conservatively estimate $L_{\rm J} / L_{\rm X}
\geq 0.05$ for Cyg X-1 and GX 339-4 at an accretion luminosity of
$L_{\rm X} \sim 10^{-2}$. Fender et al. (2001) estimated that, at an
accretion luminosity of $L_{\rm X} \sim 10^{-3}$, the black hole
transient XTE J1118+480 had $L_{\rm J} / L_{\rm X} \geq 0.2$ (see also
Corbel \& Fender 2002 for an estimate for GX 339-4). Conservatively
adopting the equality for XTE J1118+480 corresponds to $A_{\rm BH}
\sim 6 \times 10^{-3}$ in Eddington units. Equivalently the relation
between total power and jet power is given by:

\begin{equation}
L_{\rm total} = A^{-2} L_{\rm J}^{2} + L_{\rm J}
\end{equation}

In the following we shall assume that $L_{\rm total}$ is proportional
to the mass accretion rate $\dot{m}$ (i.e. all the available accretion
power goes either into the X-rays or the jet). In Eddington units this
corresponds to

\begin{equation}
L_{\rm tot} = \dot{m}
\end{equation}

which is the condition of no advection of accretion energy across the
event horizon.

We can then plot the variation of $L_{\rm X}$ and $L_{\rm J}$ as
a function of mass accretion rate. These are plotted for black holes
in the top panel of Fig 1.

We note that there are two regimes, `X-ray dominated' at higher mass
accretion rates, and `jet dominated' at lower accretion rates.  In the
X-ray dominated regime, $L_{\rm X} \propto \dot{m}$ and $L_{\rm J}
\propto \dot{m}^{0.5}$.  However in the jet dominated regime $L_{\rm
X} \propto \dot{m}^2$ and $L_{\rm J} \propto \dot{m}$. The transition
between the two regimes occurs at $L_{\rm X} = A^2 \sim 4 \times
10^{-5}$ or, equivalently, $\dot{m} = 2 A^2 \sim 7
\times 10^{-5}$. The shaded region in the top panel of Fig 1 indicates
the observed range of X-ray luminosities of black hole X-ray binaries
in `quiescence' -- if our model is correct then {\em all} of these
systems are in the jet-dominated regime, with accretion rates $10^{-6}
\la \dot{m} \la 10^{-5}$, and with jet powers one to two orders of
magnitude greater than the observed X-ray luminosity.


\section{Jet-dominated states in neutron stars ?}

A major uncertainty in knowing if the arguments outlined above apply
to neutron stars is that the relation $L_{\rm radio} \propto L_{\rm
X}^b$ has not yet been measured. Migliari et al. (2003) note that the
relation seems steeper ($b > 1$) for the atoll-type X-ray binary 4U
1728-34, but this is over a small range in X-ray flux compared to that
measured for black holes. At present we must consider that this
relation remains unmeasured, due primarily to the relative faintness
of atoll-type sources in the radio band compared to black holes
(Fender \& Hendry 2000), which results from the greater `radio
loudness' of black holes (Fender \& Kuulkers 2001; Migliari et
al. 2003). However, we will make the assumption in what follows that
the same relation does indeed apply for atoll-type NS XRBs.


As already noted, the ratio of `radio loudness' between BHC and NS
XRBs is $R_{\rm radio} \sim 30$. Using equation ({\ref{jetradio}}), this
translates into a difference in jet power of a factor 10, i.e. $A_{\rm
NS} \sim 6 \times 10^{-4}$. The X-ray luminosity below which neutron
star systems would be jet-dominated is therefore $L_{\rm X} \sim
3\times 10^{-7}$. This is comparable to the lowest X-ray luminosity
measured from a neutron star in quiescence (Garcia et al. 2001)
implying that, quite unlike black holes, we may have never observed a
neutron star in a jet-dominated state. In the lower panel of Fig 1 we
plot the variation of $L_{\rm X}$ and $L_{\rm J}$ as a function of
mass accretion rate for neutron star binaries. As for the BHCs, the
shaded region indicates the observed range of `quiescent' X-ray
luminosities. 

In the absence of core/crustal emission (see below) which is decoupled
from the accretion flow on all but the longest timescales, the
observed X-ray luminosities of `quiescent' NS XRBs correspond to a
range in accretion rate of $10^{-6} \la \dot{m} \la 10^{-4}$,
overlapping with the range in $\dot{m}$ for `quiescent' black
holes. Therefore, in at least this respect, in the model presented
here the data are consistent with both NS and BHC XRBs in `quiescence'
accreting at the same rate.

\begin{figure}
\centerline{{\epsfig{file=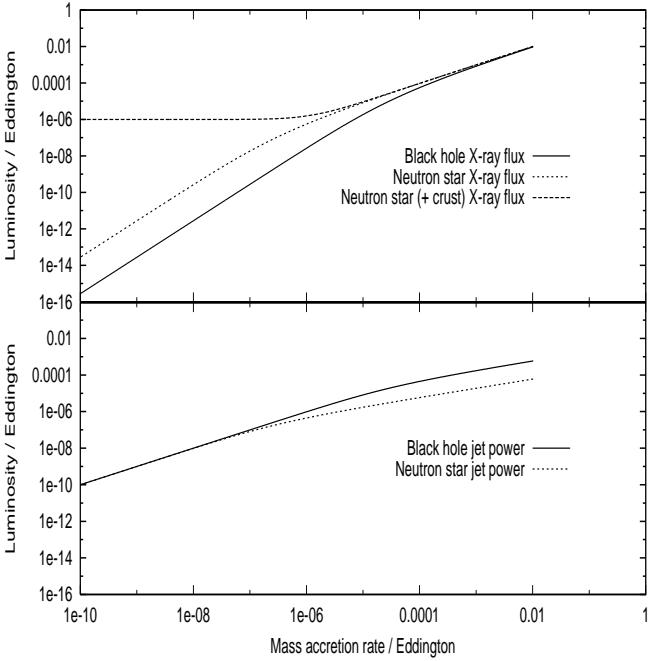, width=9cm, height=10cm, angle=0}}}
\caption{Variation of X-ray luminosity (upper panel) and jet power
(lower panel) as a function of mass accretion rate for the model
outlined in the text. Since black holes transit to the `Jet dominated'
regime at two orders of magnitude higher accretion rate than neutron
stars (Fig 1), once both classes of system are in this regime
(accretion rates corresponding to `quiescence') then neutron stars
will remain a factor of $Q_{\rm X} \sim 130$ more luminous in
X-rays. This ratio, is consistent with observations of BHCs and NS
XRBs in quiescence.  Furthermore note that at very low accretion rates
($\dot{m} \leq 10^{-6.5}$) the jet power from BHCs and NS XRBs is the
essentially the same, despite the NS XRBs being 100 times more
luminous in X-rays, and `quiescent' NS and BHC XRBs may be accreting
at the same rate.}
\end{figure}

\subsection{Core / crustal emission?}

Brown, Bildsten \& Rutledge (1998) have argued that, once accretion
has halted, neutron stars will have a luminosity in the range $5
\times 10^{32}$ -- $5 \times 10^{33}$ erg s$^{-1}$ from crustal
emission. This model seems to be supported by observations of
transient neutron star binaries in quiescence (e.g. Rutledge et
al. 2001a; Rutledge et al. 2001b; Rutledge et al. 2002; see also
Wijnands et al. 2001), although Garcia et al. (2001) argue that even
at `quiescent' levels the X-ray luminosity is dominated by accretion.
Indeed, the quiescent emission of SAX J1808.4--3658 is uncomfortably
low (5$\times10^{31} {\rm erg\,s^{-1}}$) and hard (power law index 1.5
with a blackbody contribution of less than 10 per cent) for the
neutron star crustal emission model unless the neutron star is more
massive than 1.7 M$_\odot$ (Campana et al. 2002).

If this crustal emission does exist then it adds a new term to the
total observed X-ray emission:

\begin{equation}
L_{\rm X, observed} = L_{\rm X} + L_{\rm crustal}
\end{equation}

In Fig 1 (lower panel) we also indicate the solutions with the
addition of persistent `crustal' emission to the observed X-ray flux
from a neutron star, at a level of $10^{32}$ erg s$^{-1}$,
approximately the lowest luminosity observed from a quiescent neutron
star. This has a significant effect, since this crustal luminosity, at
$\sim 10^{-6} L_{\rm Edd}$ is above that at which neutron stars would
enter the jet-dominated regime. Whereas in the case of accretion-only
luminosity, while we had not observed neutron stars in jet-dominated
regimes they were still possible, if such crustal luminosities are
ubiquitous then neutron stars will not enter the jet-dominated
regime, unless their time-averaged mass accretion rates are very low 
($\dot{m} \la 10^{-12} M_{\odot}$ $yr^{-1}$).
However, note that at the lowest accretion rates neutron stars
will make just as powerful jets as BHCs (see next section).

\section{Discussion}

This work leads naturally to some interesting consequences if
correct. We outline these below.

\subsection{X-ray luminosity as a function of mass accretion rate}

We have seen in the above that below a certain mass accretion rate BHC
X-ray binaries probably enter a jet-dominated state.  Because of their
higher `radio loudness', black holes make the transition to this
jet-dominated state at a higher mass accretion rate than neutron stars
(by a factor $(A_{\rm BH} / A_{\rm NS})^2$). Consequently, if there
are no other effects, once both NS and BH are in the jet-dominated
regime, the NS systems will be a factor $(A_{\rm BH} / A_{\rm NS})^2$
brighter in X-rays than the BH systems. Since $(A_{\rm BH} / A_{\rm
NS}) \sim 10$ then we expect a ratio of $\sim 100$ between quiescent
X-ray luminosities at the same accretion rate.  In fact, while the
expressions and plots given so far are specifically for the condition
in equation (3) and the estimated values of $A_{\rm BH}$, $A_{\rm
  NS}$, there is a more general expression for the ratio of
X-ray luminosities when both classes of object are in the
jet-dominated regime:

\begin{equation}
Q_{\rm X} = (L_{\rm X})_{NS} / (L_{\rm X})_{\rm BH} = R_{\rm radio}^{1/b}
\end{equation}

Since $R_{\rm radio} \sim 30$ and $\beta \sim 0.7$, we expect a ratio
of X-ray luminosities in quiescence of $\sim 130$, when both BHCs and
NS XRBs are in the jet-dominated regime, and at the same mass
accretion rate. This is consistent with what is observed.

The X-ray luminosities as a function of $\dot{m}$ are illustrated in
Fig 2 (top panel). As already noted, the quiescent NS XRBs may not be
quite in the jet-dominated regime; however, the BHC XRBs are clearly
in this regime, and the difference in X-ray luminosities at the same
accretion rate is already one order of magnitude at $\dot{m} \sim
10^{-5}$, increasing to $Q_{\rm X}$ at
$\dot{m} \sim 10^{-6}$ (Fig 2, top panel). We therefore find that the
observed $L_{\rm radio} \propto L_X^{0.7}$ scaling, combined with the
order of magnitude greater radio loudness of BHC XRBs, naturally
results in a signficant difference in the
quiescent luminosities of NS and BHC XRBs, as observed. 

More precisely, we expect there to be three regimes in which the ratio
of X-ray luminosities, $R_{\rm X} = (L_{\rm X})_{\rm BH} / (L_{\rm
X})_{\rm NS}$ has different values:

\vspace*{2mm}
\begin{tabular}{llcc}
(a) $(L_{\rm X})_{\rm BH} \geq A_{\rm BH}^2$ & $(L_{\rm X})_{\rm NS} \geq A_{\rm NS}^2$
  & $R_{\rm X} = 1$ \\
(b) $(L_{\rm X})_{\rm BH} \leq A_{\rm BH}^2$ & $(L_{\rm X})_{\rm NS} \geq A_{\rm NS}^2$
  & $1 \leq R_{\rm X} \leq Q_{\rm X}$ \\
(c) $(L_{\rm X})_{\rm BH} \leq A_{\rm BH}^2$ & $(L_{\rm X})_{\rm NS} \leq A_{\rm NS}^2$
  & $R_{\rm X} =  Q_{\rm X}$ \\
\end{tabular}
\vspace*{2mm}

where (a) corresponds to both classes of objects being `X-ray
dominated', (b) corresponds to BHCs being jet-dominated and NS not,
(c) corresponds to both classes being jet-dominated.  From this study
it appears that $Q \sim 130$, and that observed `quiescence'
corresponds to regimes (a) or (b), which consistent with the
observations without requiring any accretion energy to be advected
across an event horizon.

It is interesting to note that in the jet-dominated regime, the
scaling of X-ray luminosity with mass accretion rate, $L_{\rm X}
\propto \dot{m}^2$ is exactly the same as that predicted theoretically
by ADAF models (e.g. Narayan et al. 1997; Mahadevan 1997).

\subsection{Jets at the lowest accretion rates}

It would be a mistake to assume that the persistent difference in
X-ray luminosity will result in a difference in jet powers between NS
and BHC systems at the lowest luminosities. In fact, below an
accretion rate of $\dot{m} \sim 10^{-6.5}$ both NS and BHC systems
are putting the same amount of power into the jet (Fig 2, lower
panel), which dominates the power output of the system. The tiny
fraction of the total power released as X-rays is insignificant
whether its a BHC or a NS XRB one hundred times brighter.

It is also interesting to note that the ratio in radio loudness,
$R_{\rm radio}$ is maintained throughout this scenario, but for
somewhat different reasons in the two regimes. When `X-ray dominated',
BHCs are more `radio loud' because they match the NS XRBs in X-rays
but put out more radio power. However, at the lowest accretion rates
the radio power is the same but the X-ray luminosity of the BHCs is
lower, maintaining the ratio.

\subsection{X-ray jets: what if `hard' X-ray binaries are already
  jet-dominated ?}

It has been suggested that the hard X-ray spectra observed from
low/hard state BHCs may be in some, maybe all, cases optically thin
synchrotron emission directly from the jet (Markoff, Falcke \& Fender
2001; Markoff et al. 2003). This is at odds with the more standard
view of the hard X-ray spectrum as being dominated by thermal
Comptonisation from electrons with a temperature of $\sim 100$ keV
(e.g. Sunyaev \& Titarchuk 1980; Poutanen 1998, Zdziarski et
al. 2003). If it is the correct interpretation, how does it affect the
analysis performed here ?

Since in those models the BHCs are already completely jet dominated at
$\dot{m} = 0.01$, then $L_{\rm X} \propto \dot{m}^2$ (as $L_{\rm X}
\propto L_{\rm J}^2$ [equation (4)] and $L_{\rm J} \propto \dot{m}$
this is always the case for jet-dominated emission).  In fact it can
be shown that the same ratio of `quiescent' luminosities is achieved
as in the previous analysis, as long as the NS XRBs are {\em not}
already jet-dominated at $\dot{m} \sim 0.01$ (otherwise $L_{\rm X}$ in
both classes of objects would track each other). However, the
transition to the jet-dominated regime would occur at a higher
$\dot{m}$ (by approximately two orders of magnitude), meaning the
observed `quiescent' mass accretion rates would be considerably
higher than those indicated in Fig 1.

\section{Conclusions}

The results presented in this Letter are necessarily a subset of the
possible consequences of the empirical relations and model upon which
they are based. Of the more general results, to be expanded upon in a
further work, only one is given, namely that once both BHC and NS XRB
are in jet dominated states the ratio of X-ray luminosities depends
only upon two quantities which have already been measured, i.e. $b$
and $R_{\rm radio}$:

\setcounter{equation}{8}
\begin{equation}
Q_{\rm X} = R_{\rm radio}^{1/b} \sim (30)^{1/0.7} \sim 130
\end{equation}

We suggest that based upon existing observational data, `quiescent'
BHCs are in the `jet-dominated' regime and that NS XRBs are, if not
jet-dominated, close to the transition to this regime. Specifically,
if a similar value of $b$ holds for NS XRBs as for BHCs (and this is
the key observational uncertainty) then quiescent NS XRBs are, in the
most conservative case, putting $\ga 10$\% of their power into
jets. Thus the observed ratio of X-ray luminosities should be close to
$Q_{\rm X}$, consistent with what is observed.  An additional core /
crustal contribution to the X-ray emission from NS XRBs will only
widen the discrepancy.  Essentially, we find that the difference in
quiescent X-ray luminosities between NS and BHC XRBs can be mostly, if
not entirely, explained by a difference in the efficiency of jet
production between the two types of sources (the origin of which
remains unclear).

Therefore we find that the observed difference in quiescent
luminosities of neutron star and black hole candidate XRBs {\em does
not require the presence of black hole event horizons}. This should
not be taken as a statement to the effect that we do not believe that
black hole candidates contain black holes (see related discussion in
Abramowicz et al. 2003).  Certainly there are differences between the
neutron star and black hole candidate XRBs, which may be naturally
explained by the fact that black hole candidates do in fact contain
black holes. However, since there is no requirement for significant
energy to be advected into the black hole in order to explain the
ratio of quiescent X-ray luminosities, these luminosities cannot in
turn be taken as `proof' of the existence of event horizons.

Furthermore, even if, as seems probable, the BHCs do contain black
holes with event horizons, this work shows that there is no
requirement for the advection of any significant quantity of
accretion energy across the horizon. Rather, the relative radiative
inefficiency of BHCs compared to NS XRBs at low $\dot{m}$ is due to
the low radiative efficiency of the {\em jets} they are powering, not
the accretion flow itself. 

\section*{Acknowledgements}

RPF would like to thank Sera Markoff, Sebastian Heinz and Lars
Bildsten for useful discussions. PGJ would like to thank Gordon
Ogilvie for useful discussions.

\end{document}